\documentclass[twocolumn,showpacs,prb,aps,amsmath,amssymb,superscriptaddress]{revtex4}

\usepackage[dvipdfmx]{graphicx}
\usepackage{bm}
\usepackage[dvipdfmx]{color}

\usepackage{txfonts}

\begin{document}

\title{Exact Ground States of Frustrated Quantum Spin Systems Consisting of Spin-Dimer Units}

\author{Toshiya Hikihara}
\affiliation{Faculty of Science and Technology, Gunma University, Kiryu, Gunma 376-8515, Japan}
\author{Takashi Tonegawa}
\affiliation{Professor Emeritus, Kobe University, Kobe 657-8501, Japan}
\affiliation{Department of Physical Science, Osaka Prefecture University, Sakai, Osaka 599-8531, Japan}
\author{Kiyomi Okamoto}
\affiliation{College of Engineering, Shibaura Institute of Technology, Saitama 337-8570, Japan}
\author{T{\^o}ru Sakai}
\affiliation{Graduate School of Material Science, University of Hyogo, Hyogo 678-1297, Japan}
\affiliation{National Institutes for Quantum and Radiological Science and Technology (QST), SPring-8, Hyogo 679-5148, Japan}



\begin{abstract}
We study frustrated quantum spin systems consisting of dimers of spin-1/2 spins.
We derive several models that have the exact ground state of the form of the direct product of dimer states.
The ground states realized include the product state of dimer singlets, that of dimer triplets with zero magnetization, those of dimer-spin-nematic states, and those of a mixture of the dimer states.
Pseudo spin-1/2 operators emerging in each dimer are also introduced.
\end{abstract}


\maketitle

\section{Introduction}


Frustrated magnetism has been one of the central issues in condensed-matter physics for several decades.
In frustrated magnets, a competition among interactions leads to a massive degeneracy in the ground-state manifold and provides a good opportunity for perturbations such as the quantum fluctuation to realize an unconventional ground state.
Studies searching for such an unconventional ground state in frustrated quantum magnets have been performed intensively and succeeded in identifying exotic ground states, {\it e.g.}, the quantum spin-liquid in a kagome antiferromagnet\cite{YanHW2011,MeiCHW2016,Liao2016,He2016}, the vector-chirality state in the quantum magnets in a zigzag ladder\cite{NersesyanGE1998,KaburagiKH1999,HikiharaKKT2000,HikiharaKK2001,KolezhukV2005,McCulloch2008,Okunishi2008,HikiharaMFK2010}, and the spin-multipolar state in low-dimensional frustrated ferromagnets\cite{ShannonMS2006,MomoiSS2006,HikiharaKMF2008,SudanLL2009}.


Despite the efforts made for many years, studying frustrated quantum magnets is still a challenging task.
This is mainly because many powerful theoretical tools for investigating quantum spin systems are not applicable to the problem.
For instance, the mean-field approximation is not justified in investigating unconventional states without a classical long-range order.
The quantum Monte-Carlo method breaks down when applied to frustrated systems because of the notorious negative-sign problem.
Therefore, accurate results, especially exact ones, for the frustrated quantum magnets are highly desirable.


A famous example of frustrated quantum spin models with the exact ground state is the Majumdar--Ghosh model\cite{MajumdarG1969,Majumdar1970}.
The model has the form of a zigzag spin ladder and is constructed as a sum of projection operators.
It was then shown that the model has the product states of singlet pairs of nearest-neighboring spins as the ground states with a finite excitation gap.
Exact ground states with the form of the product of local-spin-unit states have also been reported for various frustrated models\cite{ShastryS1981,Xian1995,HoneckerMT2000,TakanoKS1996,SchmidtL2010,MoritaS2016,NakanoT1995,NakanoT1997,Schmidt2005,TsukanoT1997,TonegawaOHS2016}.


\begin{figure}
\includegraphics[width=80mm]{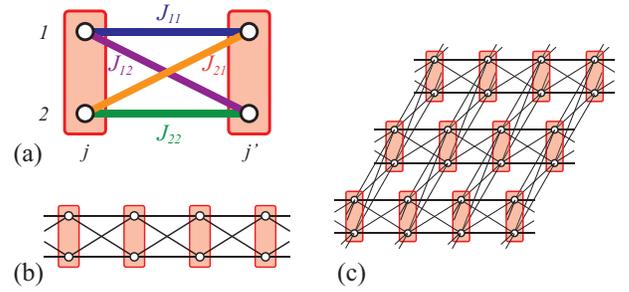}
\caption{(Color online)
(a) Schematic picture of interdimer exchange couplings.
The rectangles and circles represent dimer units and spin-1/2 spins, respectively.
The solid lines represent interdimer $XXZ$ exchange couplings.
(b) System in one-dimensional lattice.
(c) System in two-dimensional square lattice.
}
\label{fig:Ham}
\end{figure}

In this work, we discuss frustrated quantum spin systems that consist of dimers of spin-1/2 spins and have a product state of the dimer states as an exact ground state.
The model Hamiltonian is composed of the inter- and intradimer parts,
\begin{eqnarray}
\mathcal{H} &=& \mathcal{H}_{\rm inter} + \mathcal{H}_{\rm intra} .
\label{eq:H_whole}
\end{eqnarray}
The interdimer Hamiltonian consists of the $XXZ$ exchange couplings,
\begin{eqnarray}
\mathcal{H}_{\rm inter} &=&
\sum_{\langle j,j' \rangle} h_{\rm inter}(j,j')
\nonumber \\
&=& \sum_{\langle j,j' \rangle} \left[
J_{11} ({\bm S}_{1,j}, {\bm S}_{1,j'})_{\Delta}
+ J_{22} ({\bm S}_{2,j}, {\bm S}_{2,j'})_{\Delta} \right.
\nonumber \\
&&\left.~~~~~~+ J_{12} ({\bm S}_{1,j}, {\bm S}_{2,j'})_{\Delta}
+ J_{21} ({\bm S}_{2,j}, {\bm S}_{1,j'})_{\Delta} \right] ,
\label{eq:H_inter}
\end{eqnarray}
where ${\bm S}_{n,j}=(S^x_{n,j}, S^y_{n,j}, S^z_{n,j})$ $(n=1,2)$ are spin-1/2 operators in the $j$th dimer and $({\bm S}_{n,j}, {\bm S}_{n',j'})_{\Delta}$ represents the $XXZ$ anisotropic exchange coupling, i.e.,
\begin{eqnarray}
({\bm S}_{n,j}, {\bm S}_{n',j'})_{\Delta}
= S^x_{n,j} S^x_{n',j'} + S^y_{n,j} S^y_{n',j'}
+ \Delta S^z_{n,j} S^z_{n',j'} .
\end{eqnarray}
A schematic picture of the interdimer Hamiltonian is shown in Fig.\ \ref{fig:Ham}(a).
We note that the exchange constants $J_{11}, J_{22}, J_{12}$, and $J_{21}$ are different in general, while we consider the case where the anisotropy parameter $\Delta$ is the same for all the interdimer couplings.
The sum $\sum_{\langle j,j' \rangle}$ in Eq.\ (\ref{eq:H_inter}) is taken for all the bonds connected by the interdimer exchange couplings.
[See Figs. \ref{fig:Ham}(b) and \ref{fig:Ham}(c) for example.]
The structure of the lattice composed of the interdimer exchanges is basically arbitrary; 
a major part of our results is valid for any lattice in any dimension, while for some results, we require that the lattice is bipartite.

For the intradimer Hamiltonian, we consider the $XXZ$ and further anisotropic exchange couplings\cite{OnodaT2011}, the Dzyaloshinskii--Moriya (DM) coupling with the DM vector in the $z$-direction, and the Zeeman terms of uniform and staggered fields.
The intradimer Hamiltonian is then given by
\begin{eqnarray}
\mathcal{H}_{\rm intra} &=&
\sum_{j} h_{\rm intra}(j)
\nonumber \\
&=& \sum_{j} \left[ h^{XXZ}(j) + h^{\rm ani}(j) + h^{{\rm DM}z}(j) \right.
\nonumber \\
&&~~~~~~~~~~~~\left. + h^{\rm uni}(j) + h^{\rm stg} (j) \right] ,
\label{eq:H_intra} \\
h^{XXZ}(j) &=& J_d(j) ({\bm S}_{1,j}, {\bm S}_{2,j})_{\Delta_d(j)} ,
\label{eq:H_intra_XXZ} \\
h^{\rm ani}(j) &=& K_d(j) \left[ e^{i\eta_d(j)} S^+_{1,j} S^+_{2,j} + e^{-i\eta_d(j)} S^-_{1,j} S^-_{2,j} \right] ,
\label{eq:H_intra_XYZ} \\
h^{{\rm DM}z}(j) &=& D_d(j) ({\bm S}_{1,j} \times {\bm S}_{2,j})^z ,
\label{eq:H_intra_DMz} \\
h^{\rm uni}(j) &=& - H^{\rm uni}_d(j) (S^z_{1,j} + S^z_{2,j}) ,
\label{eq:H_intra_unifield} \\
h^{\rm stg}(j) &=& - H^{\rm stg}_d(j) (S^z_{1,j} - S^z_{2,j}) ,
\label{eq:H_intra_stgfield}
\end{eqnarray}
where $S^{\pm}_{n,j}=S^x_{n,j} \pm i S^y_{n,j}$.
The parameters in the intradimer Hamiltonian can change depending on the dimer position $j$ in general, but we will mainly discuss the case where they are uniform or the case where they respectively take one of two values depending on the sublattice, to which the dimer belongs, in a bipartite lattice.


We prove that the model (\ref{eq:H_whole}) in certain parameter regimes has an exact ground state, which is described by a direct product of dimer states [see Eqs.\ (\ref{eq:psi_j}) and (\ref{eq:phi_j}) for the definitions of the dimer states].
The dimer states can be the superposition of $\{ |\uparrow \downarrow\rangle_j, |\downarrow \uparrow\rangle_j \}$ or that of $\{ |\uparrow \uparrow\rangle_j, |\downarrow \downarrow\rangle_j \}$, where $| \sigma_1 \sigma_2 \rangle_j$ represents the dimer state in which the state of the spin ${\bm S}_{n,j}$ is $\sigma_n=\uparrow, \downarrow$~ $(n=1,2)$;
the dimer states include the spin-singlet state of the two spins in the dimer unit, the triplet state with zero magnetization, and the spin-nematic states.
The key idea in the proof is to show that the product state considered is an eigenstate of the interdimer Hamiltonian with zero eigenvalue\cite{TsukanoT1997}.
Then, if the product state is simultaneously the lowest-energy eigenstate of the intradimer Hamiltonian and the energy gap to the first excited states is sufficiently large, the product state, which is not disturbed by the interdimer Hamiltonian, remains the ground state even when the interdimer Hamiltonian is added to the intradimer Hamiltonian.
In such a manner, we derive several models and their exact ground states.


The paper is organized as follows.
In Sect.\ \ref{sec:pseudospin}, we introduce pseudo spin-1/2 operators composed of the two spin-1/2 spins in a dimer unit, which are used in the following argument.
The product states considered are also defined.
Our results of the exact ground state are presented in Sect.\ \ref{sec:exact}.
We propose a scheme to construct the interdimer Hamiltonian with the product states as eigenstates with zero eigenvalue in Sect.\ \ref{subsec:interdimerH}, and then discuss several examples of the model with the exact ground state in Sects.\ \ref{subsec:GSI} - \ref{subsec:shortsummary}.
Section\ \ref{sec:summary} contains a summary of our results.

\section{Pseudospin Operators and Product States}\label{sec:pseudospin}

In this section, we introduce two pseudo spin-1/2 operators defined in each dimer unit, which are used in the following discussion.
The wave functions that have the form of the direct product of dimer states and will be considered as candidates of the exact ground state are also defined.

Let us focus on the spins ${\bm S}_{1,j}$ and ${\bm S}_{2,j}$ in a dimer unit.
We construct two operators ${\bm T}_{1,j}$ and ${\bm T}_{2,j}$ defined by
\begin{eqnarray}
&&T^z_{1,j}=\frac{1}{2}\left( S^z_{1,j} - S^z_{2,j} \right) ,~~~ 
T^\pm_{1,j} = S^\pm_{1,j} S^\mp_{2,j} ,
\label{eq:T1} \\
&&T^z_{2,j}=\frac{1}{2}\left( S^z_{1,j} + S^z_{2,j} \right) ,~~~ 
T^\pm_{2,j} = S^\pm_{1,j} S^\pm_{2,j} .
\label{eq:T2}
\end{eqnarray}
One can easily find that these operators obey the following commutation relations,
\begin{eqnarray}
\left[ T^{\alpha}_{1,j} , T^{\beta}_{1,j} \right] &=& i \epsilon^{\alpha\beta\gamma} T^\gamma_{1,j} ,
\label{eq:Commu_T1} \\
\left[ T^{\alpha}_{2,j} , T^{\beta}_{2,j} \right] &=& i \epsilon^{\alpha\beta\gamma} T^\gamma_{2,j} ,
\label{eq:Commu_T2} \\
\left[ T^{\alpha}_{1,j} , T^{\beta}_{2,j} \right] &=& 0 ,
\label{eq:Commu_T1T2}
\end{eqnarray}
where $\alpha, \beta$, and $\gamma$ are $x, y$, or $z$, $T^x_{n,j}=(T^+_{n,j} + T^-_{n,j})/2, T^y_{n,j}=(T^+_{n,j} - T^-_{n,j})/(2i)$, and $\epsilon^{\alpha\beta\gamma}$ is the Levi-Civita symbol.
The operators ${\bm T}_{1,j}$ and ${\bm T}_{2,j}$ thus satisfy the usual commutation relations of spin operators and commute with each other.
Furthermore, actions of the operator ${\bm T}_{1,j}$ on the dimer states are found as
\begin{eqnarray}
&&T^z_{1,j} | \uparrow \downarrow \rangle_j 
= \frac{1}{2} | \uparrow \downarrow \rangle_j ,~~
T^z_{1,j} | \downarrow \uparrow \rangle_j 
= -\frac{1}{2} | \downarrow \uparrow \rangle_j ,
\nonumber \\
&&T^+_{1,j} | \uparrow \downarrow \rangle_j = 0 ,~~
T^+_{1,j} | \downarrow \uparrow \rangle_j = | \uparrow \downarrow \rangle_j ,
\nonumber \\
&&T^-_{1,j} | \uparrow \downarrow \rangle_j = | \downarrow \uparrow \rangle_j ,~~
T^-_{1,j} | \downarrow \uparrow \rangle_j =0 ,
\nonumber \\
&&T^\alpha_{1,j} | \uparrow \uparrow \rangle_j = T^\alpha_{1,j} | \downarrow \downarrow \rangle_j = 0 .~~~~(\alpha =x,y,z)
\label{eq:T1_action}
\end{eqnarray}
Therefore, ${\bm T}_{1,j}$ behaves as a pseudo spin-1/2 operator in the subspace $\{ | \uparrow \downarrow \rangle_j, | \downarrow \uparrow \rangle_j \}$.
The states $| \uparrow \downarrow \rangle_j$ and $| \downarrow \uparrow \rangle_j$ correspond to the states $| T^z_{1,j}=1/2 \rangle_j$ and $| T^z_{1,j}=-1/2 \rangle_j$, respectively.
In the subspace $\{ | \uparrow \uparrow \rangle_j, | \downarrow \downarrow \rangle_j \}$, ${\bm T}_{1,j}$ is zero.
Similarly, it is found that ${\bm T}_{2,j}$ behaves as a pseudo spin-1/2 operator in the subspace $\{ | \uparrow \uparrow \rangle_j, | \downarrow \downarrow \rangle_j \}$ with $| \uparrow \uparrow \rangle_j=| T^z_{2,j}=1/2 \rangle_j$ and 
$| \downarrow \downarrow \rangle_j=| T^z_{2,j}=-1/2 \rangle_j$, while ${\bm T}_{2,j}$ is zero in the subspace $\{ | \uparrow \downarrow \rangle_j, | \downarrow \uparrow \rangle_j \}$.
We note that the construction of the operators ${\bm T}_{1,j}$ and ${\bm T}_{2,j}$ from the original spin operators ${\bm S}_{1,j}$ and ${\bm S}_{2,j}$ has the same structure as that of the spin and $\eta$ operators from fermion operators\cite{Yang1989,EsslerKS1992}.

Next, we consider a unitary transformation for dimer states,
\begin{eqnarray}
U_j(\theta_j, \chi_j; \varphi_j, \zeta_j) = U_{1,j}(\theta_j, \chi_j) U_{2,j}(\varphi_j, \zeta_j) ,
\label{eq:U12}
\end{eqnarray}
with
\begin{eqnarray}
U_{1,j}(\theta_j, \chi_j) &=& \exp\left(-i \chi_j T^z_{1,j}\right) \exp\left(-i \theta_j T^y_{1,j}\right) , 
\label{eq:U1} \\
U_{2,j}(\varphi_j, \zeta_j) &=& \exp\left(-i \zeta_j T^z_{2,j}\right) \exp\left(-i \varphi_j T^y_{2,j}\right) ,
\end{eqnarray}
which represent the rotation of the pseudospins ${\bm T}_{1,j}$ and ${\bm T}_{2,j}$, respectively.
Note that $U_{1,j}(\theta_j, \chi_j)$ and $U_{2,j}(\varphi_j, \zeta_j)$ commute with each other and $U_{1,j}(\theta_j, \chi_j)$ [$U_{2,j}(\varphi_j, \zeta_j)$] is an identity operator in the subspace $\{ | \uparrow \uparrow \rangle_j, | \downarrow \downarrow \rangle_j \}$ [$\{ | \uparrow \downarrow \rangle_j, | \downarrow \uparrow \rangle_j \}$].
Using these transformations, we introduce the following dimer states,
\begin{eqnarray}
|\psi(\theta_j, \chi_j)\rangle_j &=& U_{1,j}(\theta_j, \chi_j) |\uparrow \downarrow\rangle_j
\nonumber \\
&=& e^{-i\frac{\chi_j}{2}} \cos\left( \frac{\theta_j}{2} \right) |\uparrow \downarrow\rangle_j + e^{i\frac{\chi_j}{2}} \sin\left( \frac{\theta_j}{2} \right) |\downarrow \uparrow\rangle_j ,
\nonumber \\
&&\label{eq:psi_j} \\
|\phi(\varphi_j, \zeta_j)\rangle_j &=& U_{2,j}(\varphi_j, \zeta_j) |\uparrow \uparrow\rangle_j
\nonumber \\
&=& e^{-i\frac{\zeta_j}{2}} \cos\left( \frac{\varphi_j}{2} \right) |\uparrow \uparrow\rangle_j + e^{i\frac{\zeta_j}{2}} \sin\left( \frac{\varphi_j}{2} \right) |\downarrow \downarrow\rangle_j .
\nonumber \\
&&\label{eq:phi_j}
\end{eqnarray}
We take the ranges of the phases as $-\pi \le \theta_j < \pi$, $0 \le \chi_j \le \pi$, $-\pi \le \varphi_j < \pi$, and $0 \le \zeta_j \le \pi$.
In terms of the pseudospin operators, $|\psi(\theta_j, \chi_j)\rangle_j$ and $|\phi(\varphi_j, \zeta_j)\rangle_j$ correspond respectively to the states of ${\bm T}_{1,j}$ and ${\bm T}_{2,j}$ pointing to the direction 
\begin{eqnarray}
(T^x_{1,j}, T^y_{1,j}, T^z_{1,j}) &=& (\sin\theta_j \cos\chi_j, \sin\theta_j \sin\chi_j, \cos\theta_j) ,
\label{eq:T1_direction} \\
(T^x_{2,j}, T^y_{2,j}, T^z_{2,j}) &=& (\sin\varphi_j \cos\zeta_j, \sin\varphi_j \sin\zeta_j, \cos\varphi_j) .
\label{eq:T2_direction}
\end{eqnarray}
We note that the four states $\{ |\psi(\theta_j, \chi_j)\rangle_j, |\psi(\theta_j+\pi, \chi_j)\rangle_j, |\phi(\varphi_j, \zeta_j)\rangle_j, |\phi(\varphi_j+\pi, \zeta_j)\rangle_j \}$ are orthogonal to each other and therefore can be used as an orthonormal basis for the dimer states.

Finally, it is instructive to note that the intradimer coupling terms Eqs.(\ref{eq:H_intra_XXZ}) - (\ref{eq:H_intra_stgfield}) are rewritten in terms of the operators ${\bm T}_{1,j}$ and ${\bm T}_{2,j}$ as
\begin{eqnarray}
h^{XXZ}(j) &=& J_d(j) T^x_{1,j} + J_d(j) \Delta_d(j) \left[ 2 \left( T^z_{2,j}\right)^2 - \frac{1}{4} \right] ,
\label{eq:H_intra_XXZ_T} \\
h^{\rm ani}(j) &=& 2 K_d(j) \left\{ \cos[\eta_d(j)] T^x_{2,j} - \sin[\eta_d(j)] T^y_{2,j} \right\} ,
\label{eq:H_intra_XYZ_T} \\
h^{{\rm DM}z}(j) &=& - D_d(j) T^y_{1,j} ,
\label{eq:H_intra_DMz_T} \\
h^{\rm uni}(j) &=& - 2 H^{\rm uni}_d(j) T^z_{2,j} ,
\label{eq:H_intra_unifield_T} \\
h^{\rm stg}(j) &=& - 2 H^{\rm stg}_d(j) T^z_{1,j} .
\label{eq:H_intra_stgfield_T}
\end{eqnarray}
The intradimer Hamiltonian (\ref{eq:H_intra}) has a block-diagonal form,
\begin{widetext}
\begin{eqnarray}
h_{\rm intra}(j) = \left( \begin{array}{cccc}
-\frac{1}{4}J_d(j) \Delta_d(j)-H^{\rm stg}_d(j) & \frac{1}{2}J_d(j)+\frac{i}{2} D_d(j) & 0 & 0 \\
\frac{1}{2}J_d(j)-\frac{i}{2} D_d(j) & -\frac{1}{4}J_d(j) \Delta_d(j)+H^{\rm stg}_d(j) & 0 & 0 \\
 0 & 0 & \frac{1}{4}J_d(j) \Delta_d(j)-H^{\rm uni}_d(j) & K_d(j) e^{i\eta_d(j)}\\
 0 & 0 & K_d(j) e^{-i\eta_d(j)} & \frac{1}{4}J_d(j) \Delta_d(j)+H^{\rm uni}_d(j)
\end{array}\right) ,
\label{eq:H_intra_matrix}
\end{eqnarray}
\end{widetext}
where the basis kets are arranged in the order of $\{ | \uparrow \downarrow\rangle_j, | \downarrow \uparrow\rangle_j, | \uparrow \uparrow\rangle_j, | \downarrow \downarrow\rangle_j \}$.
Therefore, the eigenstates of the intradimer Hamiltonian $h_{\rm intra}(j)$ for each dimer can be expressed as $\{ |\psi(\theta_{0j}, \chi_{0j})\rangle_j, |\psi(\theta_{0j}+\pi, \chi_{0j})\rangle_j, |\phi(\varphi_{0j}, \zeta_{0j})\rangle_j, |\phi(\varphi_{0j}+\pi, \zeta_{0j})\rangle_j \}$.
The phases $\theta_{0j}, \chi_{0j}$, $\varphi_{0j}$, and $\zeta_{0j}$ are determined as functions of the coupling constants in $h_{\rm intra}(j)$.
These results about the intradimer Hamiltonian $h_{\rm intra}(j)$ will be used in Sect.\ \ref{sec:exact} to obtain the exact ground state of model (\ref{eq:H_whole}).

\section{Exact Ground State}\label{sec:exact}

In this section, we show our main result that model (\ref{eq:H_whole}) in some parameter regions has an exact ground state of the form of the direct product of dimer states.
The strategy used to prove the result is as follows.
\begin{itemize}
\item[(i)] We first focus on the interdimer Hamiltonian (\ref{eq:H_inter}) in a certain parameter region and show that the Hamiltonian has the product states of the dimer states Eqs.\ (\ref{eq:psi_j}) and (\ref{eq:phi_j}) with some constraints on the phases $\{ \theta_j, \chi_j, \varphi_j, \zeta_j\}$ as eigenstates with zero eigenvalue.
\item[(ii)] We show that the product states with additional constraints on the phases are the eigenstates of the intradimer Hamiltonian (\ref{eq:H_intra}) considered.
At this stage, the product states obtained turn out to be eigenstates of the whole Hamiltonian (\ref{eq:H_whole}).
\item[(iii)] Finally, we specify the parameter region of the intradimer Hamiltonian, which lowers the eigenenergy of one of the eigenstates obtained in (ii) and make it be the ground state of the whole Hamiltonian.
\end{itemize}

\subsection{Interdimer Hamiltonian}\label{subsec:interdimerH}

Let us consider the interdimer exchange Hamiltonian $h_{\rm inter}(j,j')$ [Eq.\ (\ref{eq:H_inter})] between the dimers $(j,j')$.
Here, we rewrite the Hamiltonian as
\begin{eqnarray}
h_{\rm inter}(j,j') = \sum_{\epsilon, \epsilon' = \pm} 
\tilde{J}_{\epsilon \epsilon'} \left[ h^{XY}_{\epsilon \epsilon'}(j,j')
+ \Delta h^{\rm Ising}_{\epsilon \epsilon'}(j,j') \right] ,
\end{eqnarray}
with 
\begin{eqnarray}
h^{XY}_{\epsilon \epsilon'}(j,j')
&=& \left( S^x_{1,j} + \epsilon S^x_{2,j} \right) \left( S^x_{1,j'} + \epsilon' S^x_{2,j'} \right) 
\nonumber \\
&&+ \left( S^y_{1,j} + \epsilon S^y_{2,j} \right) \left( S^y_{1,j'} + \epsilon' S^y_{2,j'} \right) , 
\label{eq:h_inter_XY} \\
h^{\rm Ising}_{\epsilon \epsilon'}(j,j')
&=& \left( S^z_{1,j} + \epsilon S^z_{2,j} \right) \left( S^z_{1,j'} + \epsilon' S^z_{2,j'} \right) .
\label{eq:h_inter_Ising}
\end{eqnarray}
The original coupling constants $\{ J_{11}, J_{12}, J_{21}, J_{22}\}$ are related to $\tilde{J}_{\epsilon \epsilon'}$ as
\begin{eqnarray}
J_{11}&=&\tilde{J}_{++}+\tilde{J}_{+-}+\tilde{J}_{-+}+\tilde{J}_{--} ,
\nonumber \\
J_{12}&=&\tilde{J}_{++}-\tilde{J}_{+-}+\tilde{J}_{-+}-\tilde{J}_{--} ,
\nonumber \\
J_{21}&=&\tilde{J}_{++}+\tilde{J}_{+-}-\tilde{J}_{-+}-\tilde{J}_{--} ,
\nonumber \\
J_{22}&=&\tilde{J}_{++}-\tilde{J}_{+-}-\tilde{J}_{-+}+\tilde{J}_{--} .
\label{eq:Jnn_Jee}
\end{eqnarray}

We consider the coupling terms $h^{XY}_{\epsilon \epsilon'}(j,j')$ and $h^{\rm Ising}_{\epsilon \epsilon'}(j,j')$ acting on the following four product states of the two dimers, $|\psi(\theta_j, \chi_j)\rangle_j |\psi(\theta_{j'}, \chi_{j'})\rangle_{j'}$, $|\phi(\varphi_j, \zeta_j)\rangle_j |\phi(\varphi_{j'}, \zeta_{j'})\rangle_{j'}$, $|\psi(\theta_j, \chi_j)\rangle_j |\phi(\varphi_{j'}, \zeta_{j'})\rangle_{j'}$, and $|\phi(\varphi_j, \zeta_j)\rangle_j |\psi(\theta_{j'}, \chi_{j'})\rangle_{j'}$.
For the Ising terms, the calculation is simple since $h^{\rm Ising}_{\epsilon \epsilon'}(j,j')$'s are expressed in terms of $z$ components of the pseudospin operators such as $h^{\rm Ising}_{++}(j,j')=4 T^z_{2,j} T^z_{2,j'}$ and so on.
The operator $T^z_{1,k}$ [$T^z_{2,k}$] ($k=j,j'$) acting on the state $|\phi(\varphi_k, \zeta_k)\rangle_k$ [$|\psi(\theta_k, \chi_k)\rangle_k$] yields zero regardless of $(\varphi_k, \zeta_k)$ [$(\theta_k, \chi_k)$].
The action of the $XY$ terms $h^{XY}_{\epsilon \epsilon'}(j,j')$ on the states is slightly complicated since the terms have matrix elements between the subspaces $\{ | \uparrow \downarrow \rangle_j, | \downarrow \uparrow \rangle_j \}$ and $\{ | \uparrow \uparrow \rangle_j, | \downarrow \downarrow \rangle_j \}$.
However, we can obtain the resultant states by a straightforward calculation.
Some details of the calculation are presented in Appendix.
An interesting finding is that some of the resultant states with certain conditions on the phases $\{ \theta_j, \chi_j, \theta_{j'}, \chi_{j'} \}$ or $\{ \varphi_j, \zeta_j, \varphi_{j'}, \zeta_{j'} \}$ are zero.
The conditions on the phases required for obtaining the zero state are summarized in Table \ref{tab:zero_state}.
Using the results in the table, we can construct the interdimer Hamiltonian $\mathcal{H}_{\rm inter}$ [Eq.\ (\ref{eq:H_inter})] with exact eigenstates with zero eigenvalue.
In the following sections, we discuss some typical examples of the interdimer Hamiltonian and show that the Hamiltonian combined with an appropriate intradimer Hamiltonian has an exact ground state.

\begin{table*}
\caption{
Outcomes of the interdimer exchange couplings acting on the two-dimer product states.
The equations in the table show the condition required for having zero as the resultant state.
Here, $|s\rangle_k$ and $|t_0\rangle_k$ $(k=j,j')$ denote the dimer-singlet state [Eq.\ (\ref{eq:dimer-singlet})] and the dimer-triplet state with zero magnetization [Eq.\ (\ref{eq:dimer-triplet})], respectively.
The symbol ``0" means that the resultant state is zero regardless of the phases of the state considered, while ``-" represents the case where the resultant state is not zero for any values of the phases.
}
\label{tab:zero_state}
\begin{center}
\begin{tabular}{ccccc}
\hline
\hline
 & $|\psi(\theta_j, \chi_j)\rangle_j |\psi(\theta_{j'}, \chi_{j'})\rangle_{j'}$
 & $|\phi(\varphi_j, \zeta_j)\rangle_j |\phi(\varphi_{j'}, \zeta_{j'})\rangle_{j'}$
 & $|\psi(\theta_j, \chi_j)\rangle_j |\phi(\varphi_{j'}, \zeta_{j'})\rangle_{j'}$
 & $|\phi(\varphi_j, \zeta_j)\rangle_j |\psi(\theta_{j'}, \chi_{j'})\rangle_{j'}$ \\
\hline
$h^{XY}_{+-}(j,j')$
 & $|\psi\rangle_j=|s\rangle_j$ or $|\psi\rangle_{j'}=|t_0\rangle_{j'}$ & $\varphi_j=\varphi_{j'}$ and $\zeta_j=\zeta_{j'}$& $|\psi\rangle_j=|s\rangle_j$ & $|\psi\rangle_{j'}=|t_0\rangle_{j'}$ \\
$h^{\rm Ising}_{+-}(j,j')$
 & 0 & 0 & 0 & - \\
\hline
$h^{XY}_{-+}(j,j')$
 & $|\psi\rangle_j=|t_0\rangle_j$ or $|\psi\rangle_{j'}=|s\rangle_{j'}$ & $\varphi_j=\varphi_{j'}$ and $\zeta_j=\zeta_{j'}$& $|\psi\rangle_j=|t_0\rangle_j$ & $|\psi\rangle_{j'}=|s\rangle_{j'}$ \\
$h^{\rm Ising}_{-+}(j,j')$
 & 0 & 0 & - & 0 \\
\hline
$h^{XY}_{+-}(j,j') + h^{XY}_{-+}(j,j')$
 & $\theta_j=\theta_{j'}$ and $\chi_j=\chi_{j'}$& $\varphi_j=\varphi_{j'}$ and $\zeta_j=\zeta_{j'}$& - & - \\
$h^{\rm Ising}_{+-}(j,j') + h^{\rm Ising}_{-+}(j,j')$
 & 0 & 0 & - & - \\
\hline
$h^{XY}_{++}(j,j')$
 & $|\psi\rangle_j=|s\rangle_j$ or $|\psi\rangle_{j'}=|s\rangle_{j'}$ & $\varphi_j=-\varphi_{j'}$ and $\zeta_j=\zeta_{j'}$& $|\psi\rangle_j=|s\rangle_j$ & $|\psi\rangle_{j'}=|s\rangle_{j'}$ \\
$h^{\rm Ising}_{++}(j,j')$
 & 0 & - & 0 & 0 \\
\hline
$h^{XY}_{--}(j,j')$
 & $|\psi\rangle_j=|t_0\rangle_j$ or $|\psi\rangle_{j'}=|t_0\rangle_{j'}$ & $\varphi_j=-\varphi_{j'}$ and $\zeta_j=\zeta_{j'}$& $|\psi\rangle_j=|t_0\rangle_j$ & $|\psi\rangle_{j'}=|t_0\rangle_{j'}$ \\
$h^{\rm Ising}_{--}(j,j')$
 & - & 0 & 0 & 0 \\
\hline
$h^{XY}_{++}(j,j') + h^{XY}_{--}(j,j')$
 & $\theta_j=-\theta_{j'}$ and $\chi_j=\chi_{j'}$& $\varphi_j=-\varphi_{j'}$ and $\zeta_j=\zeta_{j'}$& - & - \\
$h^{\rm Ising}_{++}(j,j') + h^{\rm Ising}_{--}(j,j')$
 & - & - & 0 & 0 \\
\hline
\hline
\end{tabular}
\end{center}
\end{table*}

\subsection{Example I of Exact Ground States}\label{subsec:GSI}

\begin{figure}
\includegraphics[width=60mm]{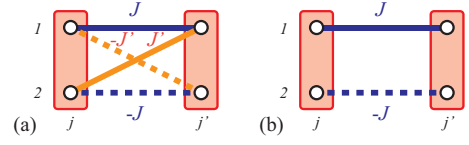}
\caption{(Color online)
Schematic pictures of (a) the interdimer Hamiltonian (\ref{eq:modelI_inter}) where $J = \tilde{J}_{+-}+\tilde{J}_{-+}$ and $J' = \tilde{J}_{+-}-\tilde{J}_{-+}$, and (b) the same Hamiltonian with $\tilde{J}_{+-}=\tilde{J}_{-+}$.
}
\label{fig:modelI}
\end{figure}

Here, we consider the interdimer Hamiltonian of the form,
\begin{eqnarray}
\mathcal{H}_{\rm inter} &=& \sum_{\langle j,j' \rangle}
\left\{ \tilde{J}_{+-} \left[ h^{XY}_{+-}(j,j')
+ \Delta h^{\rm Ising}_{+-}(j,j') \right] \right.
\nonumber \\
&&~~~~~~\left.
+ \tilde{J}_{-+} \left[ h^{XY}_{-+}(j,j')
+ \Delta h^{\rm Ising}_{-+}(j,j') \right] \right\} .
\label{eq:modelI_inter}
\end{eqnarray}
This interdimer Hamiltonian corresponds to Eq.\ (\ref{eq:H_inter}) with $J_{11}=-J_{22} = \tilde{J}_{+-}+\tilde{J}_{-+}, J_{21}=-J_{12} = \tilde{J}_{+-}-\tilde{J}_{-+}$ [see Fig.\ \ref{fig:modelI}(a)].
When $\tilde{J}_{+-}=\tilde{J}_{-+}$, the Hamiltonian is reduced to a simpler one, Eq.\ (\ref{eq:H_inter}) with $J_{11}=-J_{22}$ and $J_{21}=J_{12}=0$ [Fig.\ \ref{fig:modelI}(b)]. 

As seen in Table\ \ref{tab:zero_state}, $h^{XY}_{+-}(j,j')$ and $h^{XY}_{-+}(j,j')$ acting on $|\phi(\varphi_j, \zeta_j)\rangle_j |\phi(\varphi_{j'}, \zeta_{j'})\rangle_{j'}$ give zero for $\varphi_j = \varphi_{j'}$ and $\zeta_j=\zeta_{j'}$.
The Ising terms $h^{\rm Ising}_{+-}(j,j')$ and $h^{\rm Ising}_{-+}(j,j')$ acting on the same state also yield zero regardless of the phases.
From these results, it follows that the product state for the whole system,
\begin{eqnarray}
\prod_j |\phi(\varphi_j, \zeta_j)\rangle_j ,
\label{eq:StateI_phi}
\end{eqnarray}
is an eigenstate of model (\ref{eq:modelI_inter}) with zero eigenvalue when the phases $\varphi_j$ and $\zeta_j$ are uniform, $\varphi_j=\varphi$ and $\zeta_j=\zeta$, for arbitrary $\varphi$ and $\zeta$.
Similarly, it is found in Table\ \ref{tab:zero_state} that $h^{XY}_{+-}(j,j') + h^{XY}_{-+}(j,j')$ acting on $|\psi(\theta_j, \chi_j)\rangle_j |\psi(\theta_{j'}, \chi_{j'})\rangle_{j'}$ with $\theta_j=\theta_{j'}$ and $\chi_j=\chi_{j'}$ as well as $h^{\rm Ising}_{+-}(j,j') + h^{\rm Ising}_{-+}(j,j')$ acting on the same state (with arbitrary $\{ \theta_j, \chi_j, \theta_{j'}, \chi_{j'} \}$) yield zero.
Therefore, if the relation $\tilde{J}_{+-}=\tilde{J}_{-+}$ holds, the product state,
\begin{eqnarray}
\prod_j |\psi(\theta_j, \chi_j)\rangle_j ,
\label{eq:StateI_psi}
\end{eqnarray}
with the uniform phases $\theta_j = \theta$ and $\chi_j=\chi$ is also an eigenstate of model (\ref{eq:modelI_inter}) with zero eigenvalue for arbitrary $\theta$ and $\chi$.

Next, we include the intradimer Hamiltonian $\mathcal{H}_{\rm intra}$ [Eq.\ (\ref{eq:H_intra})] in the argument.
Here, we consider the case where the coupling constants in the intradimer Hamiltonian are uniform, i.e., $J_d(j)=J_d$, $\Delta_d(j)=\Delta_d$, and so on.
As discussed in Sect.\ \ref{sec:pseudospin}, the local intradimer Hamiltonian $h_{\rm intra}(j)$ has eigenstates $\{ |\psi(\theta_{0}, \chi_{0})\rangle_j, |\psi(\theta_{0}+\pi, \chi_{0})\rangle_j, |\phi(\varphi_{0}, \zeta_{0})\rangle_j, |\phi(\varphi_{0}+\pi, \zeta_{0})\rangle_j \}$, where the phases $\theta_0, \chi_0, \varphi_0$, and $\zeta_{0}$ are determined by the coupling constants.
The phases are independent of the position $j$ since the coupling constants are uniform.
Combining this result with the one for the interdimer Hamiltonian discussed above, we find that the two product states $\prod_j |\phi(\varphi_j, \zeta_j)\rangle_j$ with $\varphi_j = \varphi_0, \varphi_0+\pi$ and $\zeta_j=\zeta_{0}$ are eigenstates of the whole Hamiltonian $\mathcal{H}_{\rm inter} + \mathcal{H}_{\rm intra}$.
Furthermore, if $\tilde{J}_{+-}=\tilde{J}_{-+}$, the other two product states $\prod_j |\psi(\theta_j, \chi_j)\rangle_j$ with $\theta_j = \theta_0, \theta_0+\pi$ and $\chi_j=\chi_0$ are also eigenstates of the whole Hamiltonian.

We note that when the eigenstates $|\psi(\theta_{0}, \chi_{0})\rangle_j$ and  $|\psi(\theta_{0}+\pi, \chi_{0})\rangle_j$ [$|\phi(\varphi_{0}, \zeta_{0})\rangle_j$ and $|\phi(\varphi_{0}+\pi, \zeta_{0})\rangle_j$] of the local intradimer Hamiltonian $h_{\rm intra}(j)$ are degenerate, the phases $\theta_0$ and $\chi_0$ ($\varphi_0$ and $\zeta_{0}$) are not fixed.
For example, if the intradimer Hamiltonian contains only the $XXZ$ exchange term, the eigenstates $|\phi(\varphi_{0}, \zeta_{0})\rangle_j$ and $|\phi(\varphi_{0}+\pi, \zeta_{0})\rangle_j$ are degenerate regardless of the values of $J_d$ and $\Delta_d$.
If this is the case, the phases $\varphi_0$ and $\zeta_{0}$ are not fixed, and the product state $\prod_j |\phi(\varphi_0, \zeta_{0})\rangle_j$ with arbitrary $\varphi_0$ and $\zeta_{0}$ is an eigenstate of the whole Hamiltonian.
Such degenerate eigenstates were found in the model in a one-dimensional lattice\cite{TonegawaOHS2016}.

\begin{figure}
\includegraphics[width=80mm]{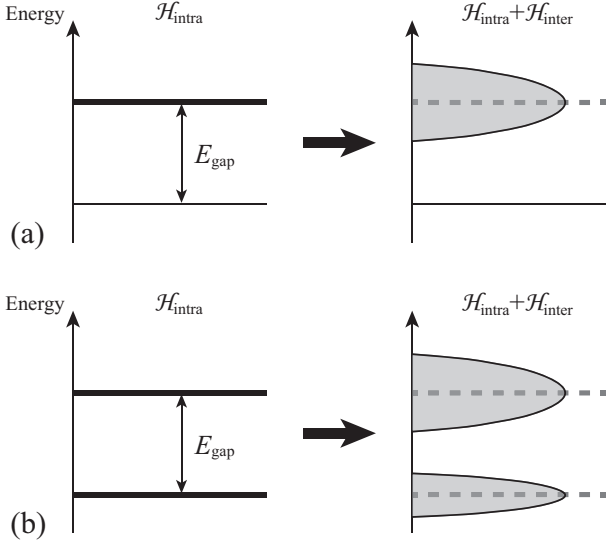}
\caption{
Schematic pictures of the density of states.
The thin and bold lines represent unique and degenerate energy levels, respectively.
(a) Case where the ground state of the intradimer Hamiltonian $\mathcal{H}_{\rm intra}$ is unique. 
When the interdimer Hamiltonian $\mathcal{H}_{\rm inter}$ is included, the ground state remains unchanged, while the first-excited states of $\mathcal{H}_{\rm intra}$, which are massively degenerate, acquire a ``kinetic energy" by $\mathcal{H}_{\rm inter}$ and form an energy band.
(b) Case where the ground states of $\mathcal{H}_{\rm intra}$ are also degenerate.
In this case, not only the manifolds of the excited states but also the manifold of the ground states form an energy band at the inclusion of $\mathcal{H}_{\rm inter}$, and the eigenstates of $\mathcal{H}_{\rm intra}$ are not the ground state of the whole Hamiltonian $\mathcal{H}_{\rm inter}+\mathcal{H}_{\rm intra}$.
}
\label{fig:mechanism}
\end{figure}

It can be proven in the following way that the eigenstates obtained above become the ground states of the whole Hamiltonian in some parameter regions.
For instance, we consider the case of the interdimer Hamiltonian (\ref{eq:modelI_inter}) with $\tilde{J}_{+-}=\tilde{J}_{-+}$ [Fig.\ \ref{fig:modelI}(b)] and the product state $\prod_j |\psi(\theta_j, \chi_j)\rangle_j$.
To prove that the state can be the ground state, it is convenient to consider the intradimer Hamiltonian first.
Let us assume that $|\psi(\theta_0, \chi_0)\rangle_j$ is the lowest-energy eigenstate of the local intradimer Hamiltonian $h_{\rm intra}(j)$ and is not degenerate to the other three eigenstates.
In this case, the ground state of the intradimer Hamiltonian $\mathcal{H}_{\rm intra}$ for the whole system is the product state $\prod_j |\psi(\theta_0, \chi_0)\rangle_j$ and there is a finite energy gap $E_{\rm gap}$ to the first excited states [see Fig.\ \ref{fig:mechanism}(a)].
The ground state is unique while the first excited states are massively degenerate ($N$-fold or more, where $N$ is the number of dimer units in the system).
When the interdimer Hamiltonian $\mathcal{H}_{\rm inter}$ is included, the ground state as well as its energy is unchanged since the state is an eigenstate of $\mathcal{H}_{\rm inter}$ with zero eigenvalue.
On the other hand, the excited states are modified by the interdimer Hamiltonian and the manifold of the first-excited states forms an energy band.
The band width should be of the order of the energy scale of the interdimer Hamiltonian.
Therefore, if the energy gap $E_{\rm gap}$ is larger than a critical value, which has the same order as the energy scale of the interdimer Hamiltonian, the product state $\prod_j |\psi(\theta_0, \chi_0)\rangle_j$ remains as the ground state of the whole Hamiltonian.
We thereby obtain the exact ground state.
We note that such an exact ground state was found in the case of a one-dimensional lattice in Refs.\ \onlinecite{TsukanoT1997} and \onlinecite{TonegawaOHS2016}.
In these studies, the Hamiltonian (\ref{eq:modelI_inter}) with $\tilde{J}_{+-}=\tilde{J}_{-+}$ [Eq.\ (\ref{eq:H_inter}) with $J_{11}=-J_{22}$ and $J_{21}=J_{12}=0$, see Fig.\ \ref{fig:modelI}(b)] was considered for the interdimer Hamiltonian, while the intradimer Hamiltonian (\ref{eq:H_intra}) is assumed to contain the $XXZ$ exchange term only [$K_d = D_d = H^{\rm uni}_d = H^{\rm stg}_d =0$].
It was found\cite{TsukanoT1997} that the model with $J_{11}=-J_{22}=1$, $\Delta=\Delta_d=1$, and $J_d > 1.134461$ has the product state of the dimer singlets,
\begin{eqnarray}
|{\rm DS} \rangle &=& \prod_j |s\rangle_j ,
\label{eq:dimer-singlet-product} \\
|s \rangle_j &=& \left| \psi\left(-\frac{\pi}{2}, 0\right) \right\rangle_j
= \frac{1}{\sqrt{2}} \left(|\uparrow \downarrow\rangle_j - |\downarrow \uparrow\rangle_j \right) ,
\label{eq:dimer-singlet}
\end{eqnarray}
as the exact ground state.
It was also reported\cite{TonegawaOHS2016} that the model with $J_{11}=-J_{22}=0.2, \Delta=1, J_d=-1$, and $0 \le \Delta_d \lesssim 0.83$ has the product state of the dimer triplets with zero magnetization,
\begin{eqnarray}
|{\rm DT}_0 \rangle &=& \prod_j |t_0\rangle_j ,
\label{eq:dimer-triplet-product} \\
|t_0 \rangle_j &=& \left| \psi\left(\frac{\pi}{2}, 0\right) \right\rangle_j
= \frac{1}{\sqrt{2}} \left(|\uparrow \downarrow\rangle_j + |\downarrow \uparrow\rangle_j \right) ,
\label{eq:dimer-triplet}
\end{eqnarray}
as the exact ground state.

Finally, we discuss the case where the local intradimer Hamiltonian $h_{\rm intra}(j)$ has degenerate ground states, choosing, as an example, the case where the states $|\phi(\varphi_{0}, \zeta_{0})\rangle_j$ and $|\phi(\varphi_{0}+\pi, \zeta_{0})\rangle_j$ are the doubly degenerate ground states of $h_{\rm intra}(j)$.
(Here, the values of $\varphi_{0}$ and $\zeta_{0}$ can be taken arbitrarily.)
This indeed occurs when $h_{\rm intra}(j)$ contains only the $XXZ$ exchange term with $J_d < 0$ and $\Delta_d > 1$ (i.e., the exchange coupling is ferromagnetic and has the Ising anisotropy).
In this case, the intradimer Hamiltonian $\mathcal{H}_{\rm intra}$ has $2^N$-fold degenerate ground states;
the Hilbert space of the ground-state manifold can be expanded by the product states $\prod_j |\phi(\varphi_j, \zeta_{0})\rangle_j$ with $\varphi_j$ taking one of the two values $\{ \varphi_0, \varphi_0+\pi \}$ arbitrarily. 
When the interdimer Hamiltonian is included, this $2^N$-fold degeneracy of the ground states is lifted:
Although two out of the degenerate ground states of $\mathcal{H}_{\rm intra}$, $\prod_j |\phi(\varphi_0, \zeta_{0})\rangle_j$ and $\prod_j |\phi(\varphi_0+\pi, \zeta_{0})\rangle_j$ (the product states with uniform $\varphi_j$), remain the eigenstates, the other states are mixed by the interdimer Hamiltonian and form the energy band  [see Fig.\ \ref{fig:mechanism}(b)].
As a result, the ground state of the whole Hamiltonian is not a simple product of dimer states but becomes a many-body entangled state.
We note that, if the anisotropic exchange coupling $h^{\rm ani}(j)$ is present in the intradimer Hamiltonian, the ground state of $\mathcal{H}_{\rm intra}$ becomes unique.
Then, if the energy gap to the first excited states is sufficiently large, a product state, 
\begin{eqnarray}
|{\rm DN}(\varphi_0, \zeta_0) \rangle = \prod_j |\phi(\varphi_0, \zeta_0)\rangle_j ,
\label{eq:dimer-spin-nematic-product}
\end{eqnarray}
where $\varphi_0$ and $\zeta_0$ are fixed according to $\mathcal{H}_{\rm intra}$, becomes the exact ground state of the whole Hamiltonian.
Since the spin-nematic correlation function in this state is long-ranged, i.e., 
\begin{eqnarray}
\langle {\rm DN}(\varphi_0, \zeta_0) | S^+_{1,j} S^+_{2,j} S^-_{1,j'} S^-_{2,j'} |{\rm DN}(\varphi_0, \zeta_0) \rangle = \frac{1}{4} \sin^2 \varphi_0 ,
\label{eq:nematic-cor-DN}
\end{eqnarray}
we call the state a spin-nematic state.
It should be noticed that this spin-nematic state is not a result of a spontaneous symmetry breaking but due to the explicit anisotropy in the Hamiltonian.
We also note that the spin-nematic state $|{\rm DN}(\varphi_0, \zeta_0) \rangle$ can be regarded as a ``ferromagnetic" state of the pseudospin ${\bm T}_{2,j}$ pointing to the direction Eq.\ (\ref{eq:T2_direction}) with $\varphi_j=\varphi_0$ and $\zeta_j=\zeta_0$.

\subsection{Example II of Exact Ground States}\label{subsec:GSII}

\begin{figure}
\includegraphics[width=80mm]{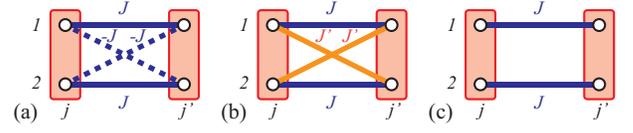}
\caption{(Color online)
Schematic pictures of (a) the interdimer Hamiltonian (\ref{eq:modelII_inter}) with $\tilde{J}_{++}=0$, where $J = \tilde{J}_{--}$, 
(b) the interdimer Hamiltonian (\ref{eq:modelII_inter}), where $J=\tilde{J}_{++}+\tilde{J}_{--}$ and $J'=\tilde{J}_{++}-\tilde{J}_{--}$, and 
(c) the same Hamiltonian as (b) with $\tilde{J}_{++}=\tilde{J}_{--}$.
For the models (b) and (c), the $XY$ case ($\Delta=0$) is considered in the text.
}
\label{fig:modelII}
\end{figure}

We discuss another example of the model with an exact ground state.
The model considered consists of the interdimer Hamiltonian of the form,
\begin{eqnarray}
\mathcal{H}_{\rm inter} &=& \sum_{\langle j,j' \rangle}
\left\{ \tilde{J}_{++} \left[ h^{XY}_{++}(j,j')
+ \Delta h^{\rm Ising}_{++}(j,j') \right] \right.
\nonumber \\
&&~~~~~~\left.
+ \tilde{J}_{--} \left[ h^{XY}_{--}(j,j')
+ \Delta h^{\rm Ising}_{--}(j,j') \right] \right\} ,
\label{eq:modelII_inter}
\end{eqnarray}
and the intradimer Hamiltonian $\mathcal{H}_{\rm intra}$ [Eq.\ (\ref{eq:H_intra})].
We consider the three parameter regions for the interdimer Hamiltonian:
\begin{itemize}
\item[(a)] $\tilde{J}_{++}=0$: In this case, the interdimer Hamiltonian is given by Eq.\ (\ref{eq:H_inter}) with $J_{11}=J_{22}=-J_{12}=-J_{21}=\tilde{J}_{--}$ [Fig.\ \ref{fig:modelII}(a)].
\item[(b)] $\Delta=0$: The exchange couplings in the interdimer Hamiltonian are of the $XY$-type and the exchange coupling constants in Eq.\ (\ref{eq:H_inter}) obey the relation $J_{11}=J_{22}=\tilde{J}_{++}+\tilde{J}_{--}$ and $J_{12}=J_{21}=\tilde{J}_{++}-\tilde{J}_{--}$ [Fig.\ \ref{fig:modelII}(b)].
\item[(c)] $\tilde{J}_{++}=\tilde{J}_{--}$ and $\Delta=0$: The interdimer exchanges are of the $XY$-type and the coupling constants in Eq.\ (\ref{eq:H_inter}) obey $J_{11}=J_{22}=\tilde{J}_{++}+\tilde{J}_{--}$ and $J_{12}=J_{21}=0$ [Fig.\ \ref{fig:modelII}(c)]. This is a special case of the model (b) above.
\end{itemize}
We also assume that the lattice is bipartite and the coupling constants in the intradimer Hamiltonian (\ref{eq:H_intra}) take one of two values depending on the sublattice A or B, i.e., $J_d(j)=J_{d,{\rm A}}$ ($j \in$ A), $J_{d,{\rm B}}$ ($j \in$ B), and so on.

We can find the parameter region of the model with an exact ground state in the same manner as described in the previous section.
From the results in Table\ \ref{tab:zero_state}, it follows that in all of the three cases listed above, the interdimer Hamiltonian has the product state,
\begin{eqnarray}
\prod_{j \in {\rm A}} |\phi(\varphi, \zeta)\rangle_j \prod_{j \in {\rm B}} |\phi(-\varphi, \zeta)\rangle_j ,
\label{eq:product-spin-nematic-stag}
\end{eqnarray}
with arbitrary $\varphi$ and $\zeta$ as an eigenstate with zero eigenvalue.
Then, if the local intradimer Hamiltonian $h_{\rm intra}(j)$ in sublattices A and B has respectively the states $|\phi(\varphi_0, \zeta_{0})\rangle_j$ and $|\phi(-\varphi_0, \zeta_{0})\rangle_j$ with certain $\varphi_0$ and $\zeta_{0}$ as an eigenstate, the product state (\ref{eq:product-spin-nematic-stag}) with $\varphi=\varphi_0$ and $\zeta=\zeta_{0}$ is an eigenstate of the whole Hamiltonian.
We note that such staggered $\varphi_j = \pm \varphi_0$ and uniform $\zeta_j = \zeta_0$ can be realized by taking the coupling constant of the anisotropic exchange terms $h^{\rm ani}(j)$ in the staggered way, $K_d(j)=K_{d,{\rm A}}<0$~ ($j \in {\rm A}$), $K_{d,{\rm B}}>0$~ ($j \in {\rm B}$) and setting $\eta_d(j)=H^{\rm uni}_d(j)=0$ in the intradimer Hamiltonian. 
In this case, the product state, 
\begin{eqnarray}
&& \prod_{j \in {\rm A}} \left| \phi\left(\frac{\pi}{2}, 0\right) \right\rangle_j \prod_{j \in {\rm B}} \left| \phi\left(-\frac{\pi}{2}, 0\right) \right\rangle_j
\nonumber \\
&& =\prod_{j \in {\rm A}} \frac{1}{\sqrt{2}} \left( |\uparrow \uparrow\rangle_j + |\downarrow \downarrow\rangle_j \right) \prod_{j \in {\rm B}} \frac{1}{\sqrt{2}} \left( |\uparrow \uparrow\rangle_j - |\downarrow \downarrow\rangle_j \right) ,
\nonumber \\
&&
\end{eqnarray}
is the eigenstate of the whole Hamiltonian.
Finally, if the product state (\ref{eq:product-spin-nematic-stag}) with $\varphi=\varphi_0$ and $\zeta=\zeta_{0}$ is the ground state of the intradimer Hamiltonian $\mathcal{H}_{\rm intra}$ with a sufficiently large excitation gap, the state becomes the exact ground state of the whole Hamiltonian $\mathcal{H}_{\rm inter}+\mathcal{H}_{\rm intra}$.
Since this ground state exhibits a staggered long-range order of the spin-nematic operator $S_{1,j}^+ S_{2,j}^+$, we call the state the antiferro-spin-nematic state.

We also see in Table\ \ref{tab:zero_state} that the interdimer Hamiltonian in case (c) mentioned above has the product state,
\begin{eqnarray}
\prod_{j \in {\rm A}} |\psi(\theta, \chi)\rangle_j \prod_{j \in {\rm B}} |\psi(-\theta, \chi)\rangle_j ,
\label{eq:product-psi-stag}
\end{eqnarray}
with arbitrary $\theta$ and $\chi$ as an eigenstate with zero eigenvalue.
Then, if the intradimer Hamiltonian has the product state (\ref{eq:product-psi-stag}) with $\theta=\theta_0$ and $\chi=\chi_{0}$ as the ground state with a sufficiently large excitation gap, the product state becomes the exact ground state of the whole Hamiltonian.
The staggered $\theta_j=\pm \theta_0$ and uniform $\chi_j=\chi_0$ of the ground state can be realized in a rather simple way: 
If the intradimer Hamiltonian contains only the $XXZ$ exchange terms with $J_d(j) = J_{d,{\rm A}} > 0$ (antiferromagnetic) for $j \in $ A and $J_d(j) = J_{d,{\rm B}} < 0$ and $0 \le \Delta_{d,{\rm B}} < 1$ (ferromagnetic and $XY$-like anisotropic) for $j \in $ B, the product state,
\begin{eqnarray}
&& \prod_{j \in {\rm A}} \left| \psi\left(-\frac{\pi}{2}, 0\right) \right\rangle_j \prod_{j \in {\rm B}} \left| \psi\left(\frac{\pi}{2}, 0\right) \right\rangle_j
\nonumber \\
&& = \prod_{j \in {\rm A}} \frac{1}{\sqrt{2}} \left( |\uparrow \downarrow\rangle_j - |\downarrow \uparrow\rangle_j \right) \prod_{j \in {\rm B}} \frac{1}{\sqrt{2}} \left( |\uparrow \downarrow\rangle_j + |\downarrow \uparrow\rangle_j \right)
\nonumber \\
&& = \prod_{j \in {\rm A}} |s\rangle_j \prod_{j \in {\rm B}} |t_0\rangle_j ,
\label{eq:product-s-t0-stag}
\end{eqnarray}
is the ground state of the intradimer Hamiltonian.
This product state becomes the exact ground state of the whole Hamiltonian if the intradimer exchange constants $J_{d,{\rm A}}$ and $|J_{d,{\rm B}}|$ are sufficiently large.
We thereby find the model with the exact ground state in which the dimer-singlet state and the dimer-triplet state with zero magnetization are arranged in a staggered fashion.
We note that when $\Delta_d(j)=0$ (i.e., not only the interdimer exchange couplings but also the intradimer ones are of the $XY$ type), the model and the ground state [Eq.\ (\ref{eq:product-s-t0-stag})] considered here are connected to the models and the states [Eqs.\ (\ref{eq:dimer-singlet-product}) and (\ref{eq:dimer-triplet-product})] discussed in the previous section through unitary transformations of the spin rotation.

\subsection{Other Examples}\label{subsec:otherGS}

In addition to the cases discussed in the preceding sections, we see in Table\ \ref{tab:zero_state} many cases where the outcome of $h^{XY}_{\epsilon \epsilon'}(j,j')$ acting on the two-dimer product state is zero.
Namely, 
\begin{eqnarray}
h^{XY}_{\epsilon \epsilon'}(j,j') | f \rangle_j | f' \rangle_{j'} = 0 ,
\label{eq:zeroeigenstate}
\end{eqnarray}
when $\epsilon~ (\epsilon') = +$ and $|f \rangle_j = | s\rangle_j$ ($|f \rangle_{j'} = | s\rangle_{j'}$), or $\epsilon~ (\epsilon') = -$ and $|f \rangle_j = | t_0\rangle_j$ ($|f \rangle_{j'} = | t_0\rangle_{j'}$).
This comes from the fact that $S^\pm_{1,j} + S^\pm_{2,j}$ ($S^\pm_{1,j} - S^\pm_{2,j}$) acting on $| s\rangle_j$ ($| t_0\rangle_j$) yields zero,
\begin{eqnarray}
(S^\pm_{1,j} + S^\pm_{2,j}) | s\rangle_j = (S^\pm_{1,j} - S^\pm_{2,j}) | t_0\rangle_j = 0 .
\label{eq:S1pmS2_zero}
\end{eqnarray}

A simple example of the application of Eq.\ (\ref{eq:zeroeigenstate}) can be found in the model consisting of the interdimer Hamiltonian,
\begin{eqnarray}
\mathcal{H}_{\rm inter} = \sum_{\langle j,j' \rangle}
\tilde{J}_{++} \left[ h^{XY}_{++}(j,j')
+ \Delta h^{\rm Ising}_{++}(j,j') \right] ,
\label{eq:modelIIIa_inter}
\end{eqnarray}
and the intradimer Hamiltonian including only the $XXZ$ exchange couplings.
For this model, the product state of the dimer-singlet states, $|{\rm DS}\rangle$ [Eq.\ (\ref{eq:dimer-singlet-product})], is the exact eigenstate of $\mathcal{H}_{\rm inter}$ with zero eigenvalue and becomes the exact ground state of the whole Hamiltonian if the $XXZ$ exchange couplings in the intradimer Hamiltonian are antiferromagnetic and sufficiently strong.
We note that this mechanism to realize the dimer-singlet-product ground state can be understood from the viewpoint that $({\bm S}_{1,j}+{\bm S}_{2,j})^2$ for each dimer unit is a good quantum number in the model.
This type of the exact ground state has been reported for various frustrated spin models\cite{Xian1995,HoneckerMT2000,TakanoKS1996,SchmidtL2010,MoritaS2016,Schmidt2005}.

Equation (\ref{eq:zeroeigenstate}) can be used to derive many other models with an exact ground state.
Let us consider, for instance, the interdimer Hamiltonian,
\begin{eqnarray}
\mathcal{H}_{\rm inter} = \sum_{\langle j,j' \rangle}
\tilde{J}_{--} \left[ h^{XY}_{--}(j,j')
+ \Delta h^{\rm Ising}_{--}(j,j') \right] ,
\label{eq:modelIIIa_inter}
\end{eqnarray}
in a bipartite lattice.
It is then found that the product state,
\begin{eqnarray}
\prod_{j \in {\rm A}} | t_0 \rangle_j \prod_{j \in {\rm B}} |\phi(\varphi_j, \zeta_j) \rangle_j ,
\end{eqnarray}
is the eigenstate of $\mathcal{H}_{\rm inter}$ with zero eigenvalue.
Therefore, if the local intradimer Hamiltonian $h_{\rm intra}(j)$ in sublattice A has $|t_0\rangle_j$ as an eigenstate and $h_{\rm intra}(j)$ in sublattice B does the spin-nematic state $|\phi(\varphi_0, \zeta_0)\rangle_j$ (with certain $\varphi_0$ and $\zeta_0$), the product state $\prod_{j \in {\rm A}} | t_0 \rangle_j \prod_{j \in {\rm B}} |\phi(\varphi_0, \zeta_0) \rangle_j$ is the eigenstate of the whole Hamiltonian.
Furthermore, if the product state is the ground state of the intradimer Hamiltonian with a sufficiently large excitation gap, the product state becomes the exact ground state of the whole Hamiltonian.
In such a manner, we can construct several models with an exact ground state written as a direct product of the dimer-singlet state $|s\rangle_j$, the dimer-triplet state with zero magnetization $|t_0\rangle_j$, and the spin-nematic state $|\phi(\varphi_0, \zeta_0)\rangle_j$.

\subsection{Short Summary of Models and Exact Ground States}\label{subsec:shortsummary}

Here, we summarize the models and their exact ground states discussed in the preceding sections.
The interdimer Hamiltonians of the models considered and the corresponding ground states are as follows:
\begin{itemize}
\item[(I)] The $XXZ$ exchange Hamiltonian (\ref{eq:modelI_inter}), which is schematically shown in Figs.\ \ref{fig:modelI}(a) and \ref{fig:modelI}(b):
For this interdimer Hamiltonian, the product of the dimer-spin-nematic states (\ref{eq:StateI_phi}) with uniform phases $\varphi_j=\varphi$ and $\zeta_j=\zeta$ is a candidate of the exact ground state of the whole Hamiltonian.

\item[(I')] The $XXZ$ exchange Hamiltonian (\ref{eq:modelI_inter}) with $\tilde{J}_{+-}=\tilde{J}_{-+}$, shown schematically in Fig.\ \ref{fig:modelI}(b):
This interdimer Hamiltonian has the product state (\ref{eq:StateI_psi}) with $\theta_j = \theta$ and $\chi_j=\chi$ as a candidate of the ground state.

\item[(II)] The $XXZ$ exchange Hamiltonian (\ref{eq:modelII_inter}) with $\tilde{J}_{++}=0$ and the $XY$ ($\Delta = 0$) exchange Hamiltonian (\ref{eq:modelII_inter}) in a bipartite lattice:
The Hamiltonians are schematically shown in Figs.\ \ref{fig:modelII}(a) - \ref{fig:modelII}(c).
For these interdimer Hamiltonians, the product of the dimer-spin-nematic states with staggered phases, Eq.\ (\ref{eq:product-spin-nematic-stag}), is a candidate of the ground state.

\item[(II')] The $XY$ ($\Delta=0$) exchange Hamiltonian (\ref{eq:modelII_inter}) with $\tilde{J}_{++}=\tilde{J}_{--}$, shown in Fig.\ \ref{fig:modelII}(c), in a bipartite lattice:
This interdimer Hamiltonian has the product state (\ref{eq:product-psi-stag}) as a candidate of the ground state.

\item[(III)] An exchange Hamiltonian in which operators acting on the $j$th dimer are expressed in terms of only $S^\alpha_{1,j} + S^\alpha_{2,j}$ or of only $S^\alpha_{1,j} - S^\alpha_{2,j}$:
With an adequate inclusion of the $XY$ and Ising terms according to Table \ref{tab:zero_state}, the interdimer Hamiltonian has a product state of the dimer singlet $|s \rangle_j$ and the dimer triplet with zero magnetization, $|t_0 \rangle_j$, with a configuration of $|s \rangle_j$ and $|t_0 \rangle_j$ corresponding to the Hamiltonian, as a candidate of the ground state.
\end{itemize}
Then, if the candidate state is a unique ground state of the intradimer Hamiltonian (\ref{eq:H_intra}) with a sufficiently large excitation gap, the state is the exact ground state of the whole system composed of the inter- and intradimer Hamiltonians, $\mathcal{H}_{\rm inter} + \mathcal{H}_{\rm intra}$.
We note that the Hamiltonian of type (I') with the dimer-singlet ground state $|{\rm DS}\rangle$ [Eq.\ (\ref{eq:dimer-singlet-product})] and the dimer-triplet ground state with zero magnetization, $|{\rm DT}_0 \rangle$ [Eq.\ (\ref{eq:dimer-triplet-product})], in a one-dimensional lattice was investigated in Refs. \onlinecite{TsukanoT1997} and \onlinecite{TonegawaOHS2016}, while the Hamiltonian of type (III), especially the one that is written in terms of $S^\alpha_{1,j} + S^\alpha_{2,j}$ and has the ground state including dimer singlets $|s \rangle_j$, was studied in the literature\cite{Xian1995,HoneckerMT2000,TakanoKS1996,SchmidtL2010,MoritaS2016,Schmidt2005}.

We emphasize that models that can be proven by our scheme to have an exact ground state are not limited to the ones mentioned above.
For instance, adding the term $\sum_{\langle j,j' \rangle} h^{\rm Ising}_{--}(j,j')$ to the interdimer Hamiltonian of type (I) above does not change the conclusion since the term acting on the ground state considered gives zero.
An inhomogeneity in coupling constants $\tilde{J}_{\epsilon \epsilon'}$ of the interdimer Hamiltonians also does not affect the conclusion.
One can thus construct a variety of models with an exact ground state using Table\ \ref{tab:zero_state}.

\section{Summary}\label{sec:summary}


In summary, we have studied frustrated quantum spin systems consisting of spin-dimer units, Eq.\ (\ref{eq:H_whole}).
We have shown that the systems in certain parameter regions have an exact ground state written in the form of the direct product of dimer states.
In the argument, we first specified the interdimer Hamiltonian which has the product state considered as an eigenstate with zero eigenvalue, and showed that the state can be the eigenstate of a certain intradimer Hamiltonian simultaneously.
We then showed that the eigenstate becomes an exact ground state of the whole Hamiltonian (the sum of the inter- and intradimer Hamiltonians) when the coupling parameters in the intradimer Hamiltonian are selected appropriately.
In such a way, we have found several models, each of which has the exact ground state of the form of the product of dimer states, including the product of the dimer-singlet states [Eq.\ (\ref{eq:dimer-singlet-product})], that of the dimer-triplet states with zero magnetization [Eq.\ (\ref{eq:dimer-triplet-product})], those of the dimer-spin-nematic states [Eq.\ (\ref{eq:dimer-spin-nematic-product})], and various products with a two-sublattice structure.
We have also introduced two operators ${\bm T}_{1,j}$ and ${\bm T}_{2,j}$ [Eqs.\ (\ref{eq:T1}) and (\ref{eq:T2})]:
The operator ${\bm T}_{1,j}$ (${\bm T}_{2,j}$) acts in the subspace $\{ |\uparrow \downarrow\rangle_j, |\downarrow \uparrow\rangle_j\}$ ($\{ |\uparrow \uparrow\rangle_j, |\downarrow \downarrow\rangle_j\}$) as a spin-1/2 operator, while it is zero in the subspace $\{ |\uparrow \uparrow\rangle_j, |\downarrow \downarrow\rangle_j\}$ ($\{ |\uparrow \downarrow\rangle_j, |\downarrow \uparrow\rangle_j\}$).

\acknowledgments
We thank Tsutomu Momoi, Kouichi Okunishi, Hiroki Nakano, and Shigeki Onoda for fruitful discussions.
T.H. was supported by JSPS KAKENHI Grant Number 15K05198.
We were also supported by JSPS KAKENHI Grant Numbers 16K05419 and 15K05882 (J-Physics) and by the Hyogo Science and Technology Association.


\appendix
\section{}

The cases where the interdimer exchange terms $h^{XY}_{\epsilon \epsilon'}(j,j')$ and $h^{\rm Ising}_{\epsilon \epsilon'}(j,j')$ acting on a two-dimer product state give zero, which are summarized in Table \ref{tab:zero_state}, can be divided into the following three groups. 

First, the Ising terms $h^{\rm Ising}_{\epsilon \epsilon'}(j,j')$ can be written in terms of $T^z_{1,k}$ and $T^z_{2,k}$ ($k=j,j'$) as mentioned in Sect. \ref{subsec:interdimerH}.
Therefore, when $h^{\rm Ising}_{\epsilon \epsilon'}(j,j')$ including $T^z_{1,k}$ [$T^z_{2,k}$] acts on $|\phi(\varphi_k, \zeta_k)\rangle_k$ [$|\psi(\theta_k, \chi_k)\rangle_k$], the outcome is zero.
These cases are listed as ``0" in Table \ref{tab:zero_state}.

Second, it follows from Eq.\ (\ref{eq:S1pmS2_zero}) that the $XY$ terms $h^{XY}_{\epsilon \epsilon'}(j,j')$ including the factor $(S^\pm_{1,k} + S^\pm_{2,k})$ [$(S^\pm_{1,k} - S^\pm_{2,k})$] yield zero when acting on the dimer-singlet state $|s\rangle_k$ [the dimer-triplet state with zero magnetization, $|t_0\rangle_k$].
These cases are listed in Table \ref{tab:zero_state} as `` $|\psi\rangle_j=|s\rangle_j$", `` $|\psi\rangle_{j}=|t_0\rangle_{j}$", and so on.

Third, there are other nontrivial cases where the $XY$ terms $h^{XY}_{\epsilon \epsilon'}(j,j')$ yield zero.
For instance, the outcome of $h^{XY}_{+-}(j,j')$ acting on $|\phi(\varphi_j, \zeta_j)\rangle_j |\phi(\varphi_{j'}, \zeta_{j'})\rangle_{j'}$ is given by
\begin{eqnarray}
&&h^{XY}_{+-}(j,j') |\phi(\varphi_j, \zeta_j)\rangle_j |\phi(\varphi_{j'}, \zeta_{j'})\rangle_{j'} 
\nonumber \\
&&= -\frac{1}{2} \left[ \cos\left( \frac{\zeta_j-\zeta_{j'}}{2}\right) \sin\left( \frac{\varphi_j-\varphi_{j'}}{2}\right) \right.
\nonumber \\
&&~~~~~~~~~\left. +i \sin\left( \frac{\zeta_j-\zeta_{j'}}{2}\right) \sin\left( \frac{\varphi_j+\varphi_{j'}}{2}\right)\right] 
\nonumber \\
&&~~~~~~~~~\times \left( | \uparrow \downarrow \rangle_j + | \downarrow \uparrow \rangle_j \right) \left( | \uparrow \downarrow \rangle_{j'} - | \downarrow \uparrow \rangle_{j'} \right) .
\label{eq:hXY+-_phiphi}
\end{eqnarray}
This resultant state becomes zero if $\varphi_j=\varphi_{j'}$ and $\zeta_j=\zeta_{j'}$.
We note that the state (\ref{eq:hXY+-_phiphi}) is zero also in the case of $\varphi_j=-\varphi_{j'}$ and $\zeta_j=\zeta_{j'}+\pi$. 
In our argument, we consider only the case of $\varphi_j=\varphi_{j'}$ and $\zeta_j=\zeta_{j'}$ since these two cases give the same state $|\phi(\varphi_j, \zeta_j)\rangle_j |\phi(\varphi_{j'}, \zeta_{j'})\rangle_{j'}$ up to an overall factor.
In addition, the state (\ref{eq:hXY+-_phiphi}) becomes zero for arbitrary $\zeta_j$ and $\zeta_{j'}$ if $\varphi_j=\varphi_{j'}=0$ or $\varphi_j=\varphi_{j'}=\pi$.
We ignore these cases of $\varphi_j=\varphi_{j'}=0$ and $\varphi_j=\varphi_{j'}=\pi$ in our argument as they correspond to the trivial states $|\phi(\varphi_j, \zeta_j)\rangle_j |\phi(\varphi_{j'}, \zeta_{j'})\rangle_{j'} = | \uparrow \uparrow \rangle_j | \uparrow \uparrow \rangle_{j'}$ and $| \downarrow \downarrow \rangle_j | \downarrow \downarrow \rangle_{j'}$, respectively.

In a similar way, one can find that the outcome is zero for the cases denoted in Table \ref{tab:zero_state} as 
``$\theta_j=\theta_{j'}$ and $\chi_j=\chi_{j'}$",
``$\varphi_j=\varphi_{j'}$ and $\zeta_j=\zeta_{j'}$", 
``$\theta_j=-\theta_{j'}$ and $\chi_j=\chi_{j'}$", and
``$\varphi_j=-\varphi_{j'}$ and $\zeta_j=\zeta_{j'}$".
The zero states obtained in these cases stem from perfect destructive interferences among the interdimer exchange processes.


\begin{thebibliography}{99}
\bibitem{YanHW2011}
S.\ Yan, D.\ A.\ Huse, and S.\ R.\ White,
Science {\bf 332}, 1173 (2011).

\bibitem{MeiCHW2016}
J.-W.\ Mei, J.-Y.\ Chen, H.\ He, and X.-G.\ Wen,
arXiv:1606.09639.

\bibitem{Liao2016}
H.\ J.\ Liao, Z.\ Y.\ Xie, J.\ Chen, Z.\ Y.\ Liu, H.\ D.\ Xie, R.\ Z.\ Huang, B.\ Normand, and T.\ Xiang,
arXiv:1610.04727.

\bibitem{He2016}
Y.-C.\ He, M.\ P.\ Zaletel, M.\ Oshikawa, and F.\ Pollmann,
arXiv:1611.06238.

\bibitem{NersesyanGE1998}
A.\ A.\ Nersesyan, A.\ O.\ Gogolin, and F.\ H.\ L.\ E\ss ler, 
Phys.\ Rev.\ Lett.\ {\bf 81}, 910 (1998).

\bibitem{KaburagiKH1999}
M.\ Kaburagi, H.\ Kawamura, and T.\ Hikihara, 
J.\ Phys.\ Soc.\ Jpn. {\bf 68}, 3185 (1999); J.\ Phys.\ Soc.\ Jpn. {\bf 83}, 128001 (2014).

\bibitem{HikiharaKKT2000}
T.\ Hikihara, M.\ Kaburagi, H.\ Kawamura, and T.\ Tonegawa,
J.\ Phys.\ Soc.\ Jpn. {\bf 69}, 259 (2000); J.\ Phys.\ Soc.\ Jpn.\ {\bf 83}, 128002 (2014).

\bibitem{HikiharaKK2001}
T.\ Hikihara, M.\ Kaburagi, and H.\ Kawamura,
Phys.\ Rev.\ B {\bf 63}, 174430 (2001); Phys.\ Rev.\ B {\bf 90}, 139906 (2014).

\bibitem{KolezhukV2005}
A.\ Kolezhuk and T.\ Vekua, 
Phys.\ Rev.\ B {\bf 72}, 094424 (2005).

\bibitem{McCulloch2008}
I.\ P.\ McCulloch, R.\ Kube, M.\ Kurz, A.\ Kleine, U.\ Schollw\"{o}ck, and A.\ K.\ Kolezhuk, 
Phys.\ Rev.\ B {\bf 77}, 094404 (2008).

\bibitem{Okunishi2008}
K.\ Okunishi,
J.\ Phys.\ Soc.\ Jpn. {\bf 77}, 114004 (2008). 

\bibitem{HikiharaMFK2010}
T.\ Hikihara, T.\ Momoi, A.\ Furusaki, and H.\ Kawamura,
Phys.\ Rev.\ B {\bf 81}, 224433 (2010).

\bibitem{ShannonMS2006}
N.\ Shannon, T.\ Momoi, and P.\ Sindzingre,
Phys.\ Rev.\ Lett.\ {\bf 96}, 027213 (2006).

\bibitem{MomoiSS2006}
T.\ Momoi, P.\ Sindzingre, and N.\ Shannon,
Phys.\ Rev.\ Lett. {\bf 97}, 257204 (2006).

\bibitem{HikiharaKMF2008}
T.\ Hikihara, L.\ Kecke, T.\ Momoi, and A.\ Furusaki,
Phys.\ Rev.\ B {\bf 78}, 144404 (2008).

\bibitem{SudanLL2009}
J.\ Sudan, A.\ L\"{u}scher, and A.\ M.\ L\"{a}uchli,
Phys.\ Rev.\ B {\bf 80}, 140402(R) (2009).


\bibitem{MajumdarG1969}
C.\ K.\ Majumdar and D.\ K.\ Ghosh, 
J.\ Math.\ Phys. {\bf 10}, 1399 (1969)

\bibitem{Majumdar1970}
C.\ K.\ Majumdar, 
J.\ Phys.\ C: Solid\ State\ Phys. {\bf 3}, 911 (1970).


\bibitem{ShastryS1981}
B.\ S.\ Shastry and B.\ Sutherland, 
Physica {\bf 108B}, 1069 (1981).

\bibitem{Xian1995}
Y.\ Xian, 
Phys.\ Rev.\ B {\bf 52}, 12485 (1995).

\bibitem{HoneckerMT2000}
A.\ Honecker, F.\ Mila, and M.\ Troyer,
Eur.\ Phys.\ J.\ B {\bf 15}, 227 (2000).

\bibitem{TakanoKS1996}
K.\ Takano, K.\ Kubo, and H.\ Sakamoto, 
J.\ Phys.:Condens.\ Matter {\bf 8}, 6405 (1996).

\bibitem{SchmidtL2010}
K.\ P.\ Schmidt and M.\ Laad, 
Phys.\ Rev.\ Lett.\ {\bf 104}, 237201 (2010).

\bibitem{MoritaS2016}
K.\ Morita and N.\ Shibata, 
J.\ Phys.\ Soc.\ Jpn.\ {\bf 85}, 033705 (2016).

\bibitem{NakanoT1995}
H. Nakano and M. Takahashi, 
J.\ Phys.\ Soc.\ Jpn. {\bf 64}, 2762 (1995).

\bibitem{NakanoT1997}
H. Nakano and M. Takahashi, 
J.\ Phys.\ Soc.\ Jpn. {\bf 66}, 228 (1997).

\bibitem{Schmidt2005}
H.-J.\ Schmidt,
J.\ Phys.\ A:\ Math.\ Gen. {\bf 38}, 2123 (2005).

\bibitem{TsukanoT1997}
M.\ Tsukano and M.\ Takahashi, 
J.\ Phys.\ Soc.\ Jpn.\ {\bf 66}, 1153 (1997).

\bibitem{TonegawaOHS2016}
T.\ Tonegawa, K.\ Okamoto, T.\ Hikihara, and T.\ Sakai, 
arXiv:1608.02064, submitted to J. Phys.: Conf. Series.

\bibitem{OnodaT2011}
S.\ Onoda and Y.\ Tanaka,
Phys.\ Rev.\ B {\bf 83}, 094411 (2011).

\bibitem{Yang1989}
C.\ N.\ Yang, 
Phys.\ Rev.\ Lett.\ {\bf 63}, 2144 (1989).

\bibitem{EsslerKS1992}
F.\ H.\ L.\ Essler, V.\ E.\ Korepin, and K.\ Schoutens,
Phys.\ Rev.\ Lett.\ {\bf 68}, 2960 (1992).





\end{thebibliography}
\end{document}